\documentclass{iaus}
\usepackage{graphicx}
\usepackage{natbib}
\graphicspath{{./fig/}}
\usepackage{bm}

\newcommand{\BoldVec}[1]{\mathchoice%
  {\mbox{\boldmath $\displaystyle     #1$}}%
  {\mbox{\boldmath $\textstyle        #1$}}%
  {\mbox{\boldmath $\scriptstyle      #1$}}%
  {\mbox{\boldmath $\scriptscriptstyle#1$}}%
}
\newcommand{\EQ}{\begin{equation}}
\newcommand{\EN}{\end{equation}}
\newcommand{\EQA}{\begin{eqnarray}}
\newcommand{\ENA}{\end{eqnarray}}

\newcommand{\mean}[1]{\overline #1}

\newcommand{\meanAA}{\overline{\mbox{\boldmath $A$}}}
\newcommand{\meanBB}{\overline{\bm{B}}}
\newcommand{\meanUU}{\overline{\bm{U}}}

\newcommand{\meanemf}{\overline{\mbox{\boldmath ${\cal E}$}} {}} 

\newcommand{\calFF}{\overline{\mbox{\boldmath ${\cal F}$}} {}}

\newcommand{\kf}{\mbox{$k_{\rm f}$}}

\newcommand{\uu}{\BoldVec{u} {}}

\newcommand{\UU}{\BoldVec{U} {}}

\newcommand{\bb}{\BoldVec{b} {}}

\newcommand{\BB}{\BoldVec{B} {}}

\newcommand{\aaa}{\BoldVec{a} {}}

\newcommand{\nab}{\BoldVec{\nabla} {}}

\def\Rm{\mbox{\rm Re}_{\rm M}}

\def\Cs{C_{\rm s}}
\def\Ca{C_{\alpha}}
\def\kf{k_{\rm f}}
\def\urms{u_{\rm rms}}

\def\etat{\eta_{\rm t}}

\def\Beq{B_{\rm eq}}
\newcommand{\ybook}[3]{, {\em #2}. #3 (#1).}

\title[Magnetic helicity fluxes and their effect on stellar dynamos]
{Magnetic helicity fluxes and their effect on stellar dynamos}

\author[Simon Candelaresi \& Axel Brandenburg]   
{Simon Candelaresi
 \and Axel Brandenburg}

\affiliation{NORDITA, AlbaNova University Center,
Roslagstullsbacken 23, SE-10691 Stockholm, Sweden; \\
and Department of Astronomy, Stockholm University,
SE 10691 Stockholm, Sweden}

\pubyear{2011}
\volume{286}  
\pagerange{464--467}
\jname{Comparative Magnetic Minima}
\editors{C. H. Mandrini, eds.}
\doi{}

\include{newcommands}

\begin{document}

\maketitle

\begin{abstract}
Magnetic helicity fluxes in turbulently driven $\alpha^{2}$ dynamos are
studied to demonstrate their ability to alleviate catastrophic quenching.
A one-dimensional mean-field formalism is used to achieve magnetic
Reynolds numbers of the order of $10^{5}$.
We study both diffusive magnetic helicity fluxes through the mid-plane
as well as those resulting from the recently proposed alternate dynamic
quenching formalism.
By adding shear we make a parameter scan for the critical values of the
shear and forcing parameters for which dynamo action occurs.
For this $\alpha\Omega$ dynamo we find that the preferred mode is
antisymmetric about the mid-plane.
This is also verified in 3-D direct numerical simulations.

\end{abstract}

\firstsection 

\section{Introduction}

The magnetic field of the Sun and other astrophysical objects, like galaxies,
show field strengths that are close to equipartition and length scales
that are much larger
than that of the underlying turbulent eddies.
Their magnetic field is assumed to be generated by a turbulent dynamo.
Heat is transformed into kinetic energy, which then generates magnetic energy,
which reaches values close
to the kinetic energy, i.e.\ they are in equipartition.
The central question in dynamo theory is under which circumstances
strong large-scale magnetic fields occur and what the mechanisms behind it are.

During the dynamo process, large- and small-scale magnetic helicities of
opposite signs are created.
The presence of small-scale helicity works against the kinetic $\alpha$-effect,
which drives the dynamo
\citep{PouquetFrischLeorat1976JFM, Brandenburg2001ApJ, FieldBlackman2002ApJ}.
As a consequence, the dynamo saturates on resistive timescales
(in the case of a periodic domain)
and to magnetic field strengths well below equipartition (in a closed domain).
This behavior becomes more pronounced
with increasing magnetic Reynolds number $\Rm$,
such that the saturation magnetic energy of the large-scale field
decreases with $\Rm^{-1}$
\citep{BS05}, for which it is called catastrophic.
Such concerns were first pointed out by \cite{VainshteinCattaneo1992}.
The quenching is particularly troublesome for astrophysical objects,
since for the Sun $\Rm = 10^{9}$ and galaxies $\Rm = 10^{18}$.

\section{Magnetic helicity fluxes}

The first part of this work addresses if fluxes of small-scale magnetic
helicity in an $\alpha^{2}$ dynamo can alleviate
the catastrophic quenching.
We want to reach as high magnetic Reynolds numbers as possible. Consequently
we consider the mean-field formalism \citep{MoffattBook1978, Radler1980book}
in one dimension, where a field
$\BB$ is split into a mean part $\meanBB$ and a fluctuating part $\bb$.
In mean-field theory the induction equation reads
\EQ \label{eq: mf induction}
\partial_{t}\meanBB =
\eta\nabla^{2}\meanBB +
\boldsymbol{\nabla}\times(\meanUU\times\meanBB +
\meanemf),
\EN
with the mean magnetic field $\meanBB$,
the mean velocity field $\meanUU$, the magnetic
diffusivity $\eta$, and the electromotive force
$\meanemf = \overline{\uu\times\bb}$, where $\uu=\UU-\meanUU$ and
$\bb=\BB-\meanBB$ are fluctuations.
A common approximation for $\meanemf$, which relates small-scale
with the large-scale fields, is
\EQ
\meanemf = \alpha\meanBB - \etat
\boldsymbol{\nabla}\times\meanBB,
\EN
where $\etat=\urms/(3\kf)$ is the turbulent magnetic diffusivity
in terms of the rms velocity $\urms$ and the wavenumber $\kf$ of the
energy-carrying eddies, and
$\alpha$ = $\alpha_{\rm K} + \alpha_{\rm M}$
is the sum of kinetic and magnetic
$\alpha$, respectively.
The kinetic $\alpha$ is the forcing term, i.e.\ the energy input to the
system.
In this model $\alpha_{\rm K}$ vanishes at the mid-plane and grows approximately
linearly with height until it rapidly falls off to $0$ at the boundary.
The magnetic $\alpha$ can be approximated by the magnetic helicity
in the fluctuating fields:
$\alpha_{\rm M} \approx \mean{h}_{\rm f}\times(\mu_0\rho_0\etat\kf^2/\Beq^2)$,
where $\mu_0$ is the vacuum permeability, $\rho_0$ is the mean density,
$\Beq=(\mu_0\rho_0)^{1/2}\urms$ is the equipartition field strength
and $\mean{h}_{\rm f} = \overline{\aaa\cdot\bb}$ the magnetic helicity in the
large-scale fields.

The advantage of this approach
is that we can use the time evolution equation
for the magnetic helicity to obtain the evolution equation for the magnetic
$\alpha$ \citep{ssHelLoss09}
\EQ \label{eq: dt alpham}
\frac{\partial \alpha_{\rm M}}{\partial t}
= -2\etat \kf^{2}
\left(
\frac{\meanemf \cdot \meanBB}{B_{\rm eq}^{2}} +
\frac{\alpha_{\rm M}}{\Rm}
\right)
-\nab\cdot\calFF_{\alpha},
\EN
where $\calFF_{\alpha}$ is the magnetic helicity flux term.
To distinguish this from the algebraic quenching
\citep{VainshteinCattaneo1992}
it is called dynamical $\alpha$-quenching.

For the flux term on the RHS of equation (\ref{eq: dt alpham})
we either choose it to be diffusive, i.e.\ $\calFF_{\alpha} =
-\kappa_{\alpha}\nab\alpha_{\rm M}$, or we take it to be proportional
to $\meanemf\times\meanAA$, where $\meanAA$ is the vector potential
of the mean field $\meanBB=\nab\times\meanAA$.
The latter expression follows from the recent realization \citep{HB11}
that terms involving $\meanemf$ should not occur in the expression for
the flux of the total magnetic helicity.
This will be referred to as the alternate quenching model.

\begin{figure}[h]
\centering
\includegraphics[width=0.65\linewidth]{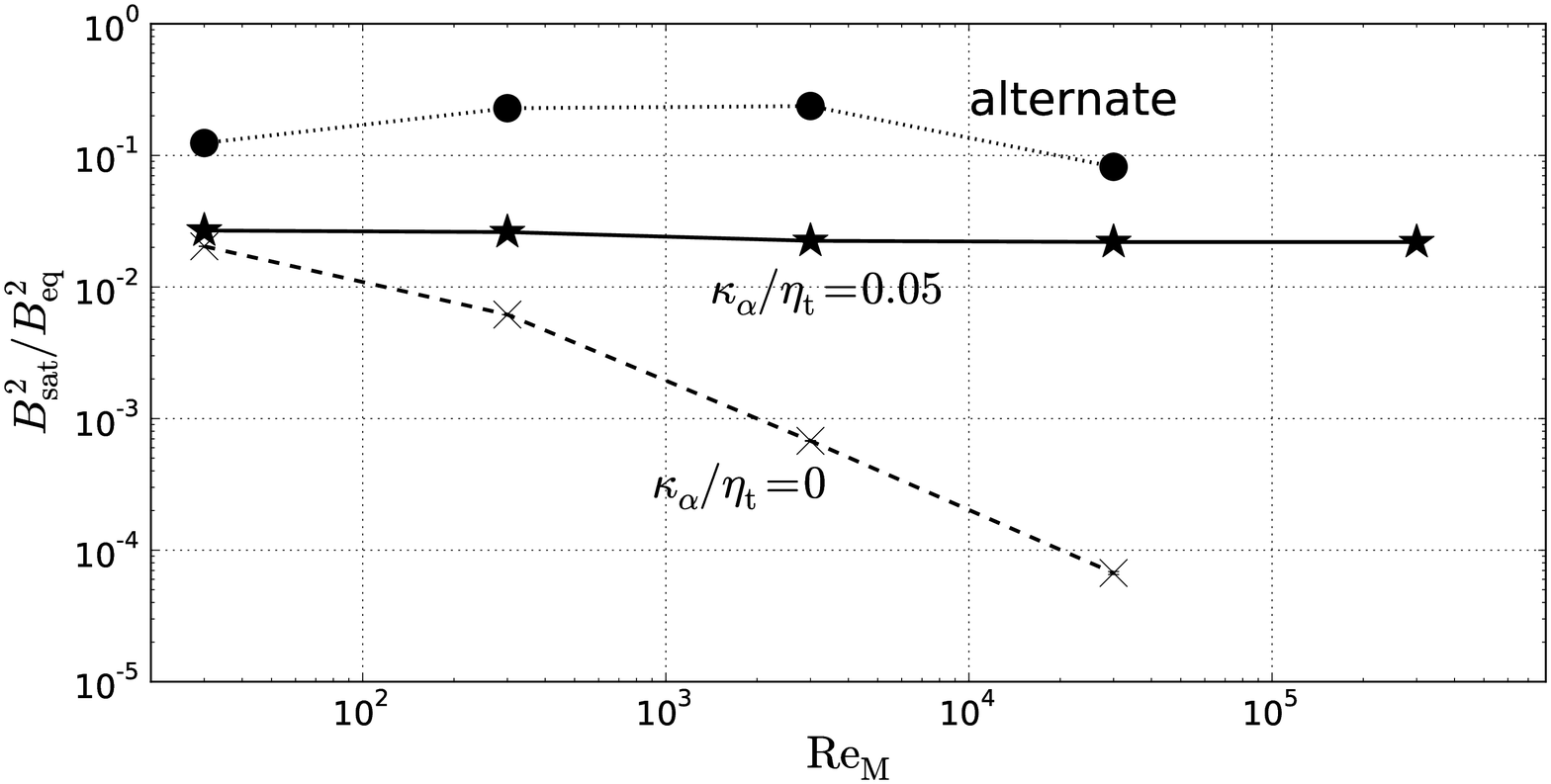}
\caption{
Saturation magnetic energy for different magnetic Reynolds numbers
with closed boundaries and diffusive fluxes (solid line) and without
(dashed line), as well as the alternate quenching formalism (dotted line).
}
\label{fig: alleviation advective}
\end{figure}

Without diffusive magnetic helicity fluxes ($\kappa_\alpha=0$),
quenching is not alleviated and the
equilibrium magnetic energy decreases as $\Rm^{-1}$
(Fig.\,\ref{fig: alleviation advective}).
We find that diffusive magnetic helicity fluxes through
the mid-plane can alleviate the
catastrophic $\alpha$ quenching and allow for magnetic field
strengths close to equipartition.
The diffusive fluxes ensure that magnetic helicity
of the small-scale field is moved from one
half of the domain to the other where it has opposite sign.
With the alternate quenching formalism we obtain larger values than with the
usual dynamical $\alpha$-quenching--even without the diffusive flux term.
The magnetic energies are however higher than expected
from simulations \citep{BS05,HB11},
which raises questions about the accuracy of the model
or its implementation.

\section{Behavior of the $\alpha\Omega$ dynamo}

In this second part we address the implications arising from adding
shear to the system and study the symmetry properties of the magnetic
field in a full domain.
The large scale velocity field in equation (\ref{eq: mf induction})
is then $\meanUU = (0, Sz, 0)$, where $S$ is the shearing amplitude
and $z$ the spatial coordinate.
We normalize the forcing amplitude $\alpha_{0}$ and the shearing amplitude
$S$ conveniently:
\EQ
C_{\alpha} = \frac{\alpha_{0}}{\eta_{\rm t}k_{1}} \qquad
C_{\rm S} = \frac{S}{\eta_{\rm t}k_{1}^{2}},
\EN
with the smallest wave vector $k_{1}$.

\begin{figure}[h]
\centering
\includegraphics[width=0.65\linewidth]{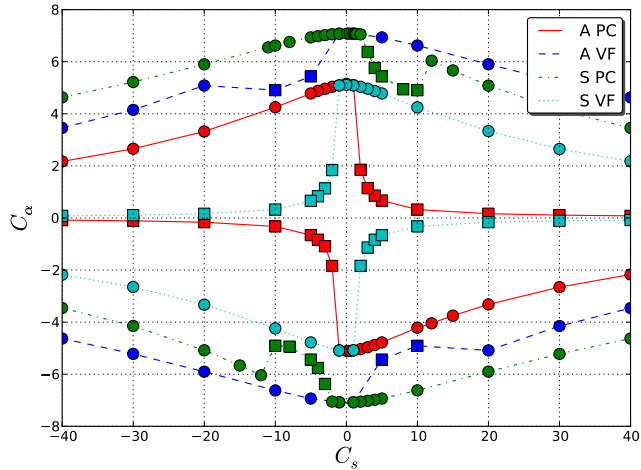}
\caption{
Critical values for the forcing amplitude $\Ca$ and the shear amplitude
$\Cs$ for an $\alpha\Omega$-dynamo in 1-D mean-field to get excited.
The circles denote oscillating solutions, while the squares denote
stationary solutions.
}
\label{fig: lambda crit}
\end{figure}

First we perform runs for the upper half of the domain using closed (perfect conductor or PC)
and open (vertical field or VF) boundaries
and impose either a symmetric or an antisymmetric mode
for the magnetic field by adjusting the boundary condition at the mid-plane.
A helical forcing is applied, which increases linearly from the mid-plane.
The critical values for the forcing and the shear parameter for which
dynamo action occurs are shown in Fig.\,\ref{fig: lambda crit}.

\begin{figure}[h]
\begin{minipage}[b]{0.48\linewidth}
\centering
\includegraphics[width=1.0\linewidth]{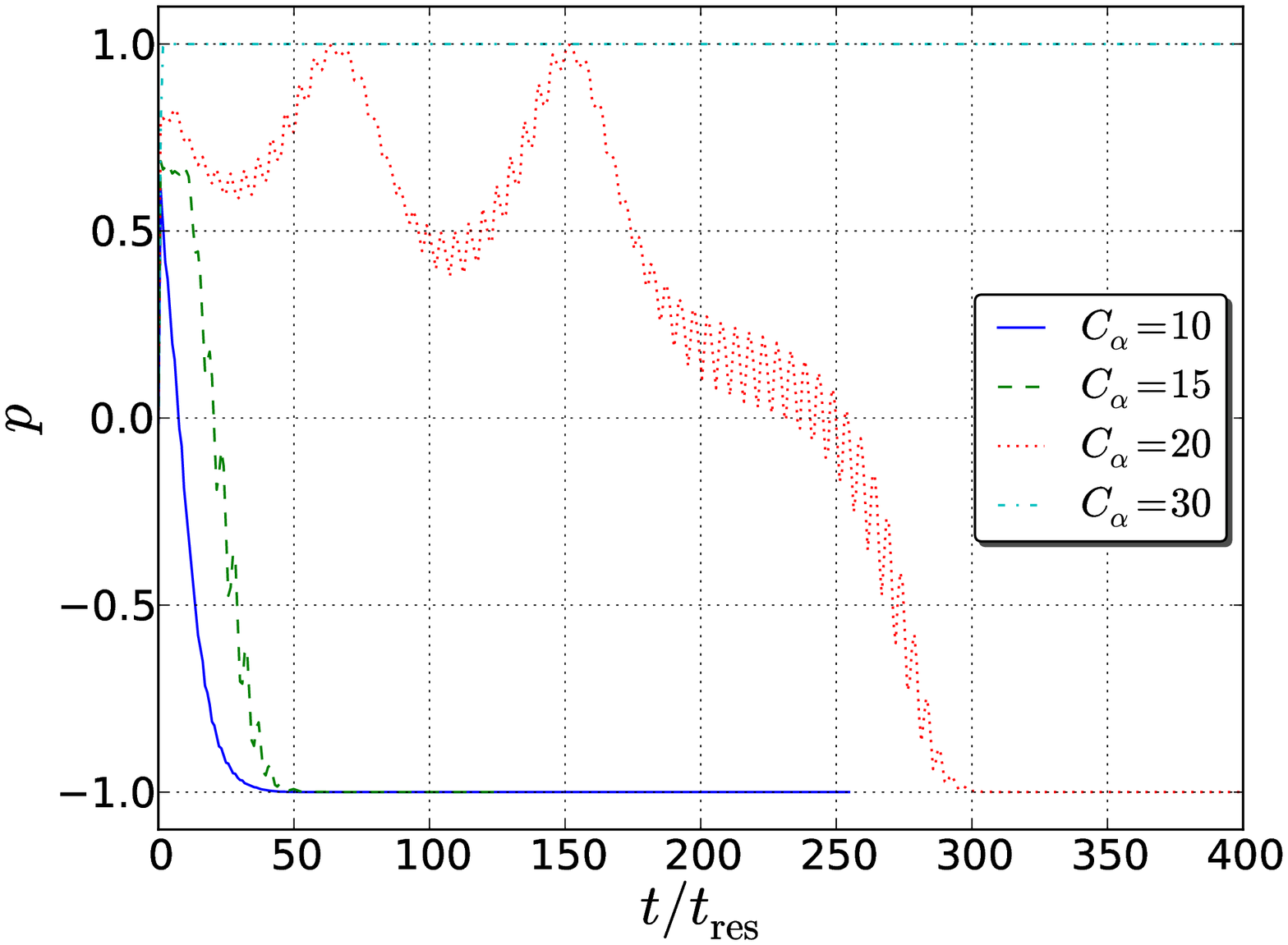}
\caption{
Parity of the magnetic field versus time for a random initial field
in 1-D mean-field.
}
\label{fig: parity MF random}
\end{minipage}
\hspace{0.04\linewidth}
\begin{minipage}[b]{0.48\linewidth}
\includegraphics[width=1.0\linewidth]{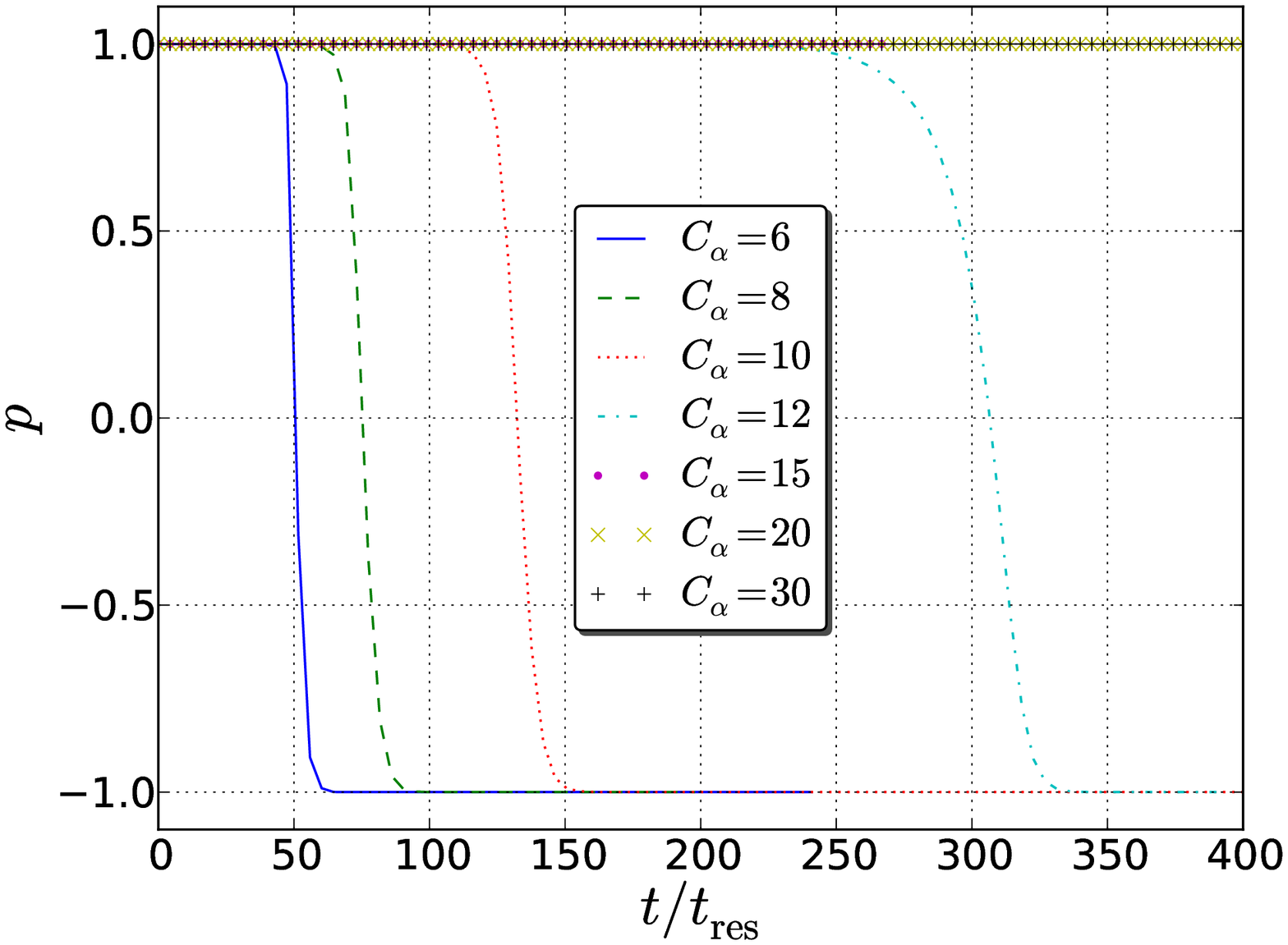}
\caption{
Parity of the magnetic field versus time for a symmetric initial field
in 1-D mean-field.
}
\label{fig: parity MF sym}
\end{minipage}
\end{figure}

Imposing the parity of the magnetic field is however unsatisfactory,
since it a priori excludes mixed modes.
Accordingly we compute the evolution of full domain systems with closed
boundaries and follow the evolution of the parity of the magnetic field.
The parity is defined such that it is 1 for a symmetric magnetic field
and $-1$ for an antisymmetric one:
\EQ
p = \frac{E_{\rm S} - E_{\rm A}}{E_{\rm S} + E_{\rm A}},
\quad
E_{\rm S/\rm A} = \int_{0}^{H} \left[\meanBB(z)\pm \meanBB(-z)\right]^{2}
\ {\rm d} z
,
\EN
with the domain height $H$.
In direct numerical simulations $B_{x}(z)$ and $B_{y}(z)$ are horizontal averages.
The field reaches an
antisymmetric solution after some resistive time
$t_{\rm res} = 1/(\eta k_{1}^{2})$ (Fig.\,\ref{fig: parity MF random}),
which depends on the forcing amplitude $\Ca$.
To check whether symmetric modes can be stable, a symmetric initial field
is imposed.
This however evolves into a symmetric field too
(Fig.\,\ref{fig: parity MF sym}), from which we conclude that it is the
stable mode.

\begin{figure}[h]
\begin{minipage}[b]{0.48\linewidth}
\centering
\includegraphics[width=1.0\linewidth]{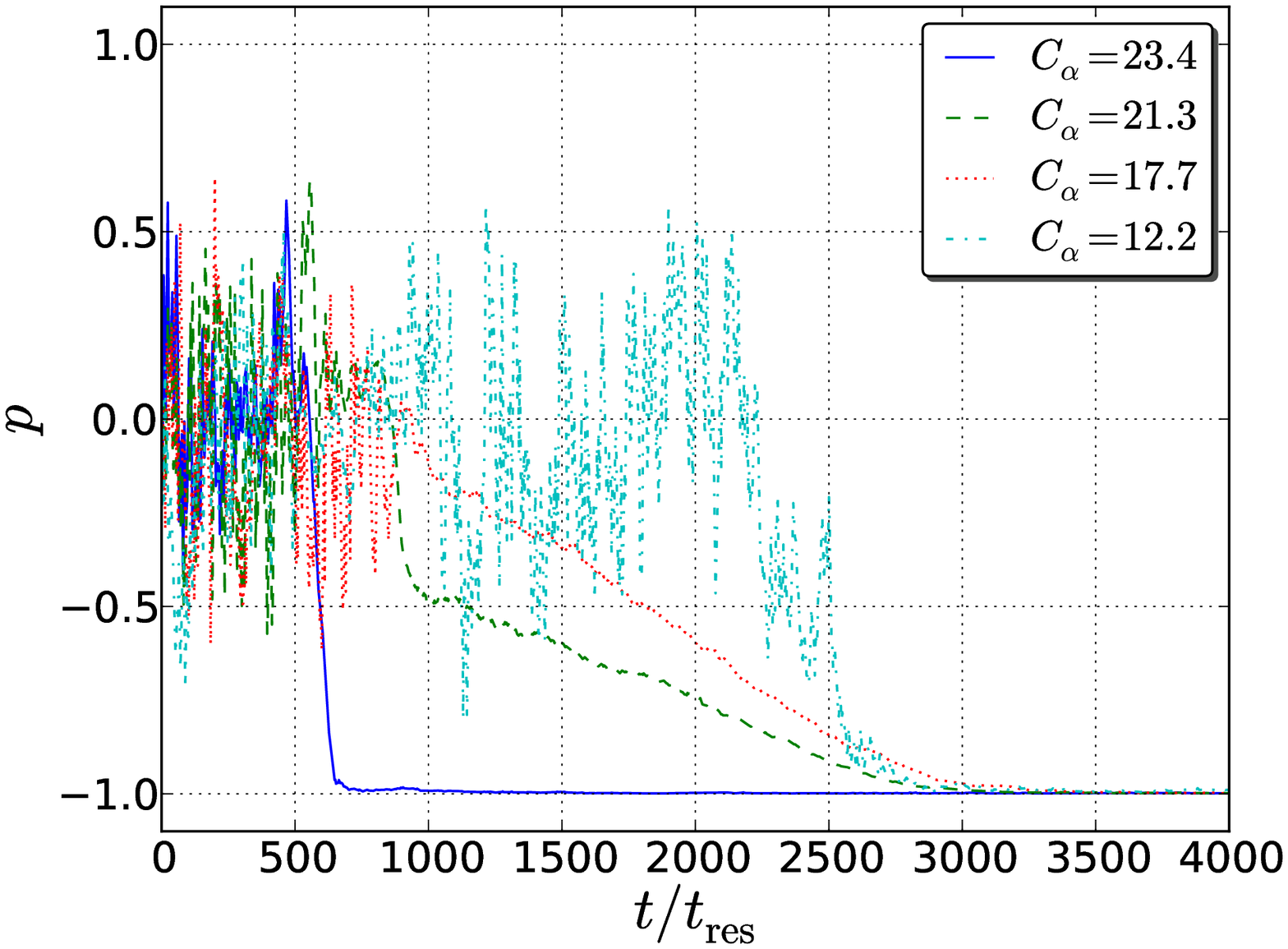}
\caption{
Parity of the magnetic field versus time for a random initial field
in 3-D DNS.
}
\label{fig: parity DNS random}
\end{minipage}
\hspace{0.04\linewidth}
\begin{minipage}[b]{0.48\linewidth}
\includegraphics[width=1.0\linewidth]{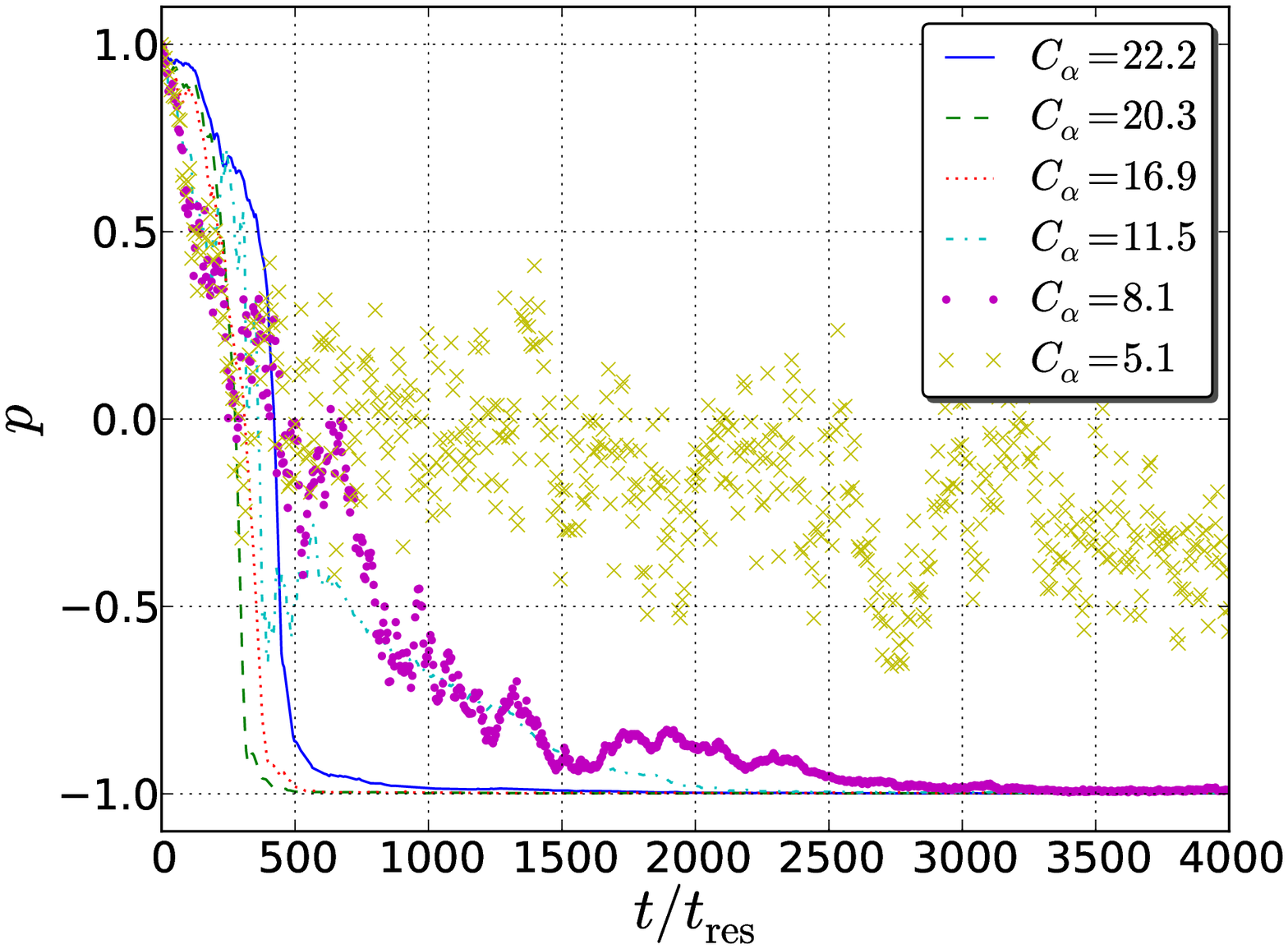}
\caption{
Parity of the magnetic field versus time for a symmetric initial field
in 3-D DNS.
}
\label{fig: parity DNS sym}
\end{minipage}
\end{figure}

The mean-field results are tested in 3-D direct numerical simulations (DNS);
Figs.\,\ref{fig: parity DNS random} and \ref{fig: parity DNS sym}.
The behavior is similar to the mean-field results.
The preferred mode is always the antisymmetric one and the time
for flipping increases with the forcing amplitude $\Ca$.
This is however very preliminary work and has to be studied in more detail.

\section{Conclusions}

The present work has shown that the magnetic helicity flux divergences
within the domain are able to alleviate catastrophic quenching.
This is also true for the fluxes implied by the alternate dynamical
quenching model of \cite{HB11}.
However, those results deserve further numerical verification.
Further, we have shown that, for the model with magnetic helicity
fluxes through the mid-plane, the preferred mode is indeed dipolar,
i.e.\ of odd parity.
Here, both mean-field models and DNS are found to be in agreement.

\begin{discussion}

\discuss{Sacha Brun}{Is there a reason that your system prefers antisymmetric
solutions? It seems linked to your choice of parameters.}

\discuss{Simon Candelaresi}{So far we do not see a reason for that. But we see a
parameter dependence of the transition time. We
will look at the growth rate of the modes independence of the
parameters. This will give us some better clue if also mixed or
symmetric modes are preferred.}

\discuss{Gustavo Guerrero}{Is there a regime where the advective flux removes
all the mean field out of the domain?}

\discuss{Simon Candelaresi}{If the advective flux is too high the magnetic
field gets shed before it is enhanced, which kills the dynamo.
So, there is a window for the advection strength for which it is
beneficial for the dynamo.}
\end{discussion}

\end{document}